# An Analysis of Factors Influencing Metro Station Ridership: Insights from Taipei Metro

Yuxin He
Department of Systems Engineering and
Engineering Management
City University of Hong Kong
Kowloon, Hong Kong
yuxinhe2-c@my.cityu.edu.hk

Yang Zhao
Centre for Systems Informatics Engineering
City University of Hong Kong
Kowloon, Hong Kong
yang.zhao@my.cityu.edu.hk

Kwok Leung Tsui
School of Data Science
City University of Hong Kong
Kowloon, Hong Kong
kltsui@cityu.edu.h

*Abstract*— Travel demand analysis at the planning stage is important for metro system development. In practice, travel demand can be affected by various factors. This paper focuses on investigating the factors influencing Taipei metro ridership at station level over varying time periods. Ordinary Least Square (OLS) multiple regression models with backward stepwise feature selection are employed to identify the influencing factors, including land use, social economic, accessibility, network structure information, etc. Network structure factors are creatively quantified based on complex network theory to accurately measure the related information. To enhance goodness-of-fit, the dummy variable distinguishing transportation hub is incorporated in the modeling. The main findings in this paper are three-fold: First, there is no distinct difference between influencing factors of boarding and those of alighting; Second, ridership is significantly associated with the number of nearby shopping malls, distance to city center, days since opening, nearby bus stations and dummy variable for transportation hub; Finally, the ridership on weekdays is mainly affected by commuting activities, while the ridership on weekends is driven by commercial access.

## I. INTRODUCTION

In transportation planning, ridership modeling and estimating is the basis for analyzing travel demand and further understanding the planning feasibility and sustainability. Metro ridership at station level is a critical element for determining the scale of stations and access facilities. As one of the best-known models, the four-step (generation, distribution, mode choice, and assignment) model has dominated the history of transport modeling since the 1950s [1]. However, the four-step model has many potential problems in practice [2], such as limitation in model accuracy, low data precision, insensitivity to land use, institutional barriers, and high expense [3]. It is generally effective for forecasting transit ridership on a regional scale rather than more detailed scales (such as station level)[4].

Direct demand models based on regression analysis is a complementary approach, which estimates ridership as a function of influence factors within the station catchment areas [3],[5]-[8]. Here, a catchment area is the geographic area for which a station attracts passengers. The size and shape of a catchment area will depend on how accessible a station is and how far it is from alternative facilities. For metro stations, one can use buffers to create circular catchment areas by a specific distance or use Thiessen polygons to illustrate the areas most accessible to each station. The advantages of direct demand models in travel analysis have been highlighted by Walters and Cervero [9] and summarized by Cardozo et al.[4] as "simplicity of use, easy interpretation of results, immediate response, and low cost".

With regard to dependent variables of regression models, the average weekday ridership was selected in most of the relevant studies, such as the research of Kuby et al.[7], Chu[8], Sohn and Shim[10], Guerra et al.[11]. In addition, monthly ridership was adopted in the research of Gutierrez et al. in 2011[3]. Zhao et al.[12]chose the annual average weekday ridership as the dependent variable of the regression model. To date, few studies analyzed average weekend ridership and compare the influencing factors on ridership of weekdays and weekends. However, the difference between travel patterns on weekdays and weekends has a significant impact on the analysis of influencing factors on ridership. An omission of travel patterns could lead to an incorrect analysis of influencing factors on ridership.

Concerning explanatory variables of regression, they can be roughly divided into the following categories, Land use, social economic, intermodal traffic access and network structure[7]. For the first kind of variables, Estupinan and Rodriguez [13], Loo et al.[14], Sohn and Shim [10], Gutierrez et al. [3], Sung and Oh[15], Choi et al.[5], Cardozo et al.[4] and Zhao et al. [12]considered commercial, residence, education, entertainment and other mixed land use as explanatory variables. For the second kind of variables, factors considered mainly are population, employment, and automobile ownership ratio. Chu [8], Kuby et al. [7], Loo et al. [14], Sohn and Shim [10], Gutierrez et al. [3], Choi et al. [5], Cardozo et al. [4], Guerra et al. [11], Thompson et al. [16], Zhao et al. [12] and Gao et al.[17] analyzed the influence of population and employment on transit ridership. Chu [8], Loo et al.[14], Thompson et al [16], Cardozo et al. [4] and Zhao et al.[12] considered the relationship between automobile ownership and the ridership. For the third category, intermodal traffic access factors, Chu[8], Kuby et al.[7], Estupinan and Rodriguez[13], Loo et al. [14], Sung

This research work was partly supported by the Research Grants Council Theme- based Research Scheme (Project No. T32-101/15-R).

and Oh [15], Gutierrez et al.[3], Choi et al.[5], Cardozo et al. [4], Guerra et al. [11] and Zhao et al. [12] analyzed the influence of bus feeder system on transit ridership. Moreover, Kuby et al. [7], Estupinan and Rodriguez [13], Loo et al.[14], Sohn and Shim [10], Thompson et al. [16], Choi et al.[5], Guerra et al. [11] and Zhao et al. [12] studied on the station accessibility. Finally, with regard to the effect of metro network structure on ridership, dealing with hypotheses relating to station spacing, interline transfer points, centrality, and so on, can produce a certain influence on ridership. Based on several practical experiences, whether the metro station is a transfer station or a terminal, and whether the station is located at an important position in the metro network could have potential effects on station ridership. Kuby et al. [7] took the influence of transfer station and terminal station on ridership into account, but both of them were regarded as dummy variables in the regression model. Sohn and Shim [10] and Thompson et al. [16] considered the factor of transfer station but which was not categorized into network structure. So far, few relevant studies have been carried out from the perspective of complex network theory, such as quantifying the transfer and terminal stations by the calculation of nodes' degree centrality and betweenness centrality of the network, which lays the foundation for quantitative analysis of the effect of network structure on ridership.

Our study was designed to assess the factors driving metro station ridership in Taipei metropolitan area in 2015. Taipei Metro is a rapid transit system serving Taipei metropolitan area. The boarding and alighting ridership data for 108 stations of Taipei metro of a whole week from Oct.12 to Oct.18 in 2015 as well as the data related to four categories of explanatory variables were collected. Among them, land use variables measured nearby sites of residence, entertainment, services, commercial, education and working. Intermodal traffic accessibility variables referred to feeder bus system. Network structure variables were related to degree centrality, betweenness centrality, and the distance from each station to city center. Finally, social economic variables consisted of days since the metro stations opening and population distribution of residents in the whole city. The purpose of the study was to quantify the effects of these factors on average weekday and weekend ridership of Taipei metro stations. The key improvements of this study over prior research are listed as follows: a) Different travel patterns at different levels of day of the week are taken into account. b) Network structures as a type of factors are quantified based on the measurements in the field of complex network. c) The data we need are less and easy to collect. Through statistical analysis, we also found influential points and add the dummy variable distinguishing transportation hub into the multiple regression model, which makes regression fit well and also perform well via cross-validation, avoiding the overfitting.

II. EMPIRICAL STUDY AREA AND DATA

This paper investigated the impacts of factors on metro ridership at station level in Taipei metropolitan area, including Taipei City, New Taipei city, Keelung city and even Taoyuan city more generally, is supported by a relatively large Metro transportation network, consisting of 5 lines and 108 stations, operating on 131.1 kilometers of revenue track. The population of Taipei city, as the area center, is about 2,695,704, the area is 272 km$^2$, and the population density is 9,918 persons/km$^2$. This density ranks Taipei as the seventh most densely populated city in the world[1].

The Taipei metro boarding and alighting ridership data used in the research are collected from the website of Taipei Rapid Transit[2]. These data cover a time span of 7 days from October 12th (Mon) to 18th (Sun) in the year of 2015. The census data were collected from the website of Worldpop[3], which only provided the raster files of population distribution in the year 2015. During the data preprocessing, the raster files were resampled to fit the cell size of metro station buffer within 500 meters.

*A. Dependent Variable*

This paper aims to identify different factors influencing the ridership at station level on weekdays and weekends. As mentioned above, the travel demands and travel patterns are different in different time periods. According to the preliminary statistics, the average daily ridership of weekdays is about 4,151,932, and the average daily ridership of weekends is about 3,782,303, which is less than that of weekdays. It indicates the daily trip frequency of weekdays is higher than weekends. Different regression models with different dependent variables, those are average weekday ridership and average weekend ridership, will be built intending to find the factors influencing the station-level ridership in different time periods.

*B. Explanatory Variables*

The explanatory variables represent factors hypothesized to influence station ridership (TableⅠ). The variables can be classified into four categories: 1) Land use variables; 2) Intermodal traffic access variables; 3) Network structure variables; and 4) Social economic variables [7].

*1)* Land use variables: Evaluating the walking distance to Metro stations is the critical first step, and the zone within the walking distance from a particular location, here refers to metro station, is defined as the pedestrian catchment area (PCA). Several scholars have carried out the research to determine the walking distance to transit stations [18]-[20], and the evaluated distance ranged from 400m to 800m. This distance is neither necessarily static from city to city nor constant as variables changing, and a great number of studies have conducted to define different spatial ranges for PCAs, such as Gutiérrez et al.[3], Choi et al.[5], and Guerra et al. [11]. However, this is not the focus of this article, for simplicity, we defined the range of PCA of Taipei metro stations as 500m, as noted by Kim et al. [21] that the average friendly walking distance was generally assumed to be 500m in large and middle-sized cities. Then, all of the land use-related data within a PCA were crawled from Google Map with the assistance of API, and land use variables consist of stations' accessible sites of residence, entertainment, services, business, education, and work. Specifically, the information covers the number of residence, hotels, schools, universities, offices, hospitals, banks, and shopping malls within PCA.

---
[1] Source: http://210.69.61.217/pxweb2007-tp/Dialog/Saveshow.asp
[2] Source: http://english.metro.taipei/ct.asp?xItem=1056489&ctNode=70217&mp=122036
[3] Source: http://www.worldpop.org.uk/data/get_data/

TABLE I. THE TABLE OF EXPLANATORY VARIABLES SUMMARY S

| Categories | Explanatory variables | Acronym | Minimum value | Average value | Maximum value |
|---|---|---|---|---|---|
| Land use | The number of residential units | *Residence* | 1 | 7.454 | 20 |
| | The number of hotels | *Hotel* | 0 | 11.01 | 153 |
| | The number of shopping malls | *Shopping* | 0 | 6.5 | 37 |
| | The number of schools | *School* | 1 | 12.28 | 45 |
| | The number of offices | *offices* | 0 | 4.222 | 14 |
| | The number of banks | *Bank* | 0 | 17.4 | 64 |
| | The number of bus stations | *Bus* | 7 | 23.95 | 45 |
| | The number of hospitals | *Hospital* | 0 | 6.861 | 37 |
| | The number of universities | *University* | 0 | 1.759 | 14 |
| Intermodal traffic accessibility | The number of bus stations | *Bus* | 7 | 23.95 | 45 |
| Network structure | Distance to the city center | *Dis_to_center* | 0.4204 | 7.2514 | 19.5844 |
| | Degree centrality | *Degree* | 0.01869 | 0.04015 | 0.07477 |
| | Betweenness centrality | *Betweenness* | 0 | 0.09772 | 0.45943 |
| Social economic | Population | *Pop* | 0.6685 | 158.7655 | 410.8630 |
| | Days since opening | *Days_open* | 26 | 4287 | 7065 |

*2) Intermodal traffic access variables:* As for intermodal traffic access, here we only considered the feeder bus system[22], and the related data indicating the number of bus stations nearby metro stations were also crawled from Google Map.

*3) Network structure variables:* In this paper, network structure variables comprised the distance to city center, which is Taipei city government, located in Hsinyi District, degree centrality, and betweenness centrality of the metro network nodes, which were correlated to the identity of stations like transfer stations and terminal stations, and the importance of stations in the aspect of centrality of the network. Previous studies (e.g., [7]) usually regards transfer stations and terminal stations as the dummy variables, but not combined with the quantified calculation measures based on the complex network theory, which contains much more information than dummy variables.

*4) Social economic variables:* With regard to social economic variables, they consist of population distribution of Taipei metropolitan area in the year of 2015 and days since metro stations opening. The census data were processed with ArcGIS 10.2. Fig. 1 showed stations distribution of Taipei metro and the population distribution of Taipei metropolitan area in the year of 2015. The information of days since metro lines and stations opening was collected from Wikipedia[4].

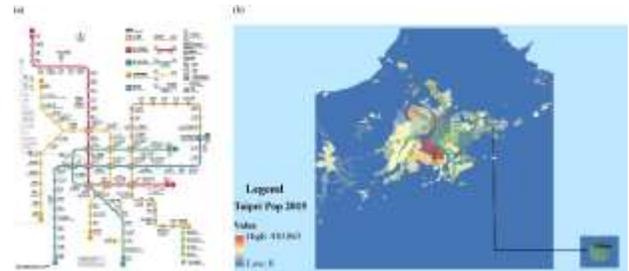

Fig. 1. Taipei metro stations distribution and population distribution. (a) Schematic route map of Taipei Metro[1]. (b) Population distribution of Taipei metropolitan area and 500 m buffers of metro stations.

Meanwhile, buffers of each metro station with a radius of 500 m were created by using ArcGIS, which was also illustrated in Fig. 1. According to Fig. 1's preliminary visualization can be noted that the population density is relatively high near the metro region. Therefore, the population data were resampled within 500 meters buffers.

### III. METHODOLOGY

To achieve the objective of the research to investigate the relationship between multiple factors and metro station ridership, we conducted two statistical analyses: stepwise regressions to select variables and multiple regression analysis.

#### A. Variables Selection

There are 14 explanatory variables shown in Table I initially, however, too many variables may cause multi-collinearity, adding noise to the estimation of other quantities that we are interested in, and even overfitting. Actually, we want to explain the dependent variable in the simplest way, so redundant explanatory variables should be removed. Therefore, before running the regression, the backward stepwise method according to Akaike Information Criterion (AIC) will be adopted to select variables and get a grip on complexity. AIC penalizes large size models and so will tend to prefer a most fitted model with the simplest parameters.

---

[4] Source: https://en.wikipedia.org/wiki/Taipei_Metro

Backward stepwise method is the simplest of all variable selection procedures. It starts with all the variables in the model, and at each step, a variable may be removed according to AIC, finally the model with the minimal AIC can be found, and variables remained are going to explain dependent variable in the regression model. Therefore, the backward stepwise method according to AIC will be implemented to the linear regression model, however, the results after stepwise regression performed not well in terms of low R-square (e.g. Multiple R-squared of the regression model for the average weekday ridership is 0.5789, and the regression model for the average weekend ridership is 0.5534).

*B. Improve the goodness-of-fit*

The results calculated by the backward regression procedure showed that the goodness-of-fit of each model is not good enough, so we seek ways to improve the goodness-of-fit of the regression model.

First of all, we did outlier test and found that No.81 sample point was an influential point from the influence plot shown in Fig. 2(a). Through look back upon the sample, we found that No.81 is Taipei Main station, which is the main transportation hub for both the city and for northern Taiwan. Taipei Main Station is home to the following transportation services: Metro - Taipei MRT, Train - Taiwan Railways, Taiwan High-Speed Rail, Taiwan Taoyuan International Airport MRT. This is the reason why the average daily ridership of Taipei Main Station is much larger than that of other stations (eg: The average daily ridership of a whole week (oct.12nd-oct.18th) of Taipei Main Station is the largest among all stations, which is 300,416 (Fig. 2(b)), and the second largest one is Ximen Station, which is 139,110).

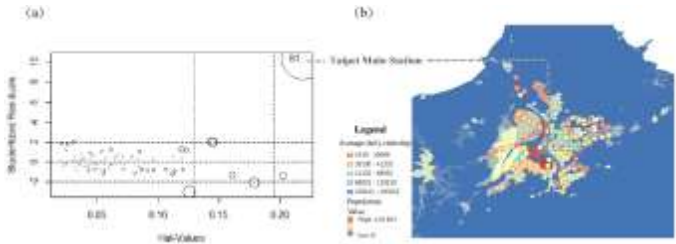

Fig. 3. Influential points in the samples of Taipei metro station ridership. (a) Regression Influence plot for the whole week average daily ridership. (b) Spatial distribution of the whole week average daily ridership and population.

In this case, we added a dummy variable for transportation hub into our multiple regression model in order to improve the performance of regression.

## IV. RESULTS AND DISCUSSION

Before running OLS multiple regression model, some variables should be transformed by a deterministic mathematical function. On the one hand, transformation can make it easier to visualize data and improve interpretability, on the other hand, if the linearity between two variables fails to hold, even approximately, it is sometimes possible to transform either the independent or dependent variables in the regression model to improve the linearity. Meanwhile, in order to observe pairwise relationships between the variables

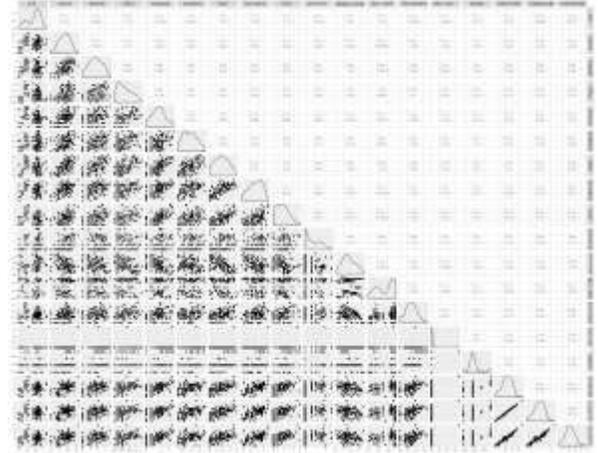

Fig. 2. Scatterplot matrix of variables.

after transformation, the scatterplot matrix is as following:

Through the transformation of raw data, we can note the pairwise relationship between two variables more clearly(see Fig. 3), and then a number of statistical analyses were run to test the 15 different variables (including the dummy variable for transportation hub) and develop the best model for explaining and predicting metro station ridership.

Initially, models were run with backward stepwise regression, which helped narrow the list of worthwhile explanatory variables. The variables that were clearly not significant in explaining either average weekday ridership or average weekend ridership were bank, hotel, school, university, residence, and hospital. All of these weren't selected in the final regression model, some of which had been expected to be positively associated with ridership in the regression model without adding the dummy variable for transportation hub.

Most of the remaining explanatory variables were highly significant in both models corresponding to different periods. And the explanatory variables of final 6 models are shown in Table Ⅱ.

TABLE II. VARIABLES INCLUDED IN THE MODELS

| Variables | Model 1 | Model 2 | Model 3 | Model 4 | Model 5 | Model 6 |
|---|---|---|---|---|---|---|
| Dependent variable | Weekday_ridership | Weekday_boarding | Weekday_alighting | Weekend_ridership | Weekend_boarding | Weekend_alighting |
| Explanatory variables | Pop | Pop | Pop | Shopping | Shopping | Shopping |
| | Office | Office | Office | Bus | Bus | Bus |
| | Shopping | Shopping | Shopping | Dis_to_center | Dis_to_center | Dis_to_center |
| | Bus | Bus | Bus | Days_open | Days_open | Days_open |
| | Dis_to | Dis_to_ | Dis_to_ | Trans_h | Trans_h | Trans_h |

|  | _center | center | center | ub | ub | ub |
|---|---|---|---|---|---|---|
|  | Days_open | Days_open | Days_open |  |  |  |
|  | Betweenness | Betweenness | Degree |  |  |  |
|  | Trans_hub | Trans_hub | Trans_hub |  |  |  |

From Table Ⅱ we can note that there is no significant difference among the explanatory variables selected in the total ridership, boardings, and alightings. Thus we can consider that there is no significant difference between the influencing factors on boarding ridership and alighting ridership. Hence, we will only discuss the influencing factors on total ridership of average weekday and average weekend. The results of Models 1 and 4 are summarized in Table Ⅲ and Table Ⅳ.

TABLE III.    RESULT OF MODEL 1

| Explanatory variables | Estimate | Standard error | t(Est/SE) | Pr(>|t|) | Significance |
|---|---|---|---|---|---|
| Intercept | -2.074e+04 | 7.608e+03 | -2.726 | 0.007589 | ** |
| Pop | 3.214e+01 | 1.794e+01 | 1.791 | 0.076295 | . |
| Office | 1.007e+03 | 5.676e+02 | 1.774 | 0.079093 | . |
| Shopping | 1.333e+03 | 2.697e+02 | 4.944 | 3.13e-06 | *** |
| Bus | 7.841e+02 | 2.190e+02 | 3.580 | 0.000535 | *** |
| Dis_to_center | 6.504e+02 | 4.061e+02 | 1.602 | 0.112426 |  |
| Days_open | 2.764e+00 | 7.103e-01 | 3.892 | 0.000180 | *** |
| Betweenness | 4.152e+04 | 2.152e+04 | 1.930 | 0.056472 | . |
| Trans_hub | 1.892e+05 | 1.762e+05 | 10.741 | < 2e-16 | *** |
| Signif. codes:  0 '***' 0.001 '**' 0.01 '*' 0.05 '.' 0.1 ' ' 1 ||||||
| *Diagnostic* ||||||
| Residual standard error | 16260 || n | 108 ||
| R-square | 0.7829 || DF | 99 ||
| Adjusted R-square | 0.7653 || F-statistic | 44.62 ||
| 10 Fold Cross-Validated R-square | 0.7179995 || P-value | < 2.2e-16 ||
| Change | 0.0649005 || AIC | 2411.477 ||

TABLE IV.    RESULT OF MODEL 4

| Explanatory variables | Estimate | Standard error | t(Est/SE) | Pr(>|t|) | Significance |
|---|---|---|---|---|---|
| Intercept | -2.278e+04 | 7.905e+03 | -2.881 | 0.00483 | ** |
| Shopping | 2.003e+03 | 2.821e+02 | 7.102 | 1.69e-10 | *** |
| Bus | 7.670e+02 | 2.314e+02 | 3.315 | 0.00127 | ** |
| Dis_to_center | 1.075e+03 | 4.317e+02 | 2.491 | 0.01435 | * |
| Days_open | 3.792e+00 | 7.239e-01 | 5.239 | 8.74e-07 | *** |
| Trans_hub | 2.539e+05 | 1.907e+04 | 13.313 | <2e-16 | *** |
| Signif. codes:  0 '***' 0.001 '**' 0.01 '*' 0.05 '.' 0.1 ' ' 1 ||||||
| *Diagnostic* ||||||
| Residual standard error | 17880 || n | 108 ||
| R-square | 0.8146 || DF | 102 ||
| Adjusted R-square | 0.8056 || F-statistic | 89.66 ||
| 10 Fold Cross-Validated R-square | 0.7620926 || P-value | < 2.2e-16 ||
| Change | 0.0525074 || AIC | 2429.228 ||

The model 1, with 99 degrees of freedom (DF), has an R-square value of 0.7829 (adjusted R-square of 0.7653), and an F-statistic value of 44.62, significant at the 0.000 level. The model thus explains 78% of the variance of average weekday ridership over all stations. This model can also get a relative high R-square via 10 fold cross-validation, 0.7180, indicating that there doesn't exist overfitting and the model generalizes to the independent dataset. In addition, model 4 for the average weekend ridership regression performs better according to the value of adjusted R square, which means that we only need to know the information of the number of nearby shopping malls, bus stations, distance to city center, days since opening, and whether it is a transportation hub of metro stations, we can use OLS multiple regression model to explain 81% of the dependent variable, average weekend ridership, and meanwhile, the data related to these explanatory variables are quite easy to obtain. The model shows statistically significant and strong explanatory power.

Moreover, the common explanatory variables of two models indicate the main factors affecting metro ridership at station level, which includes the number of shopping malls, bus stations, distance to city center, days since opening and the dummy variable for transportation hub, while different explanatory variables of different models note that different factors affect the ridership at station-level of different times. Concerning model 1, population distribution, the number of offices and betweenness centrality contribute to affect the average weekday ridership, and for model 4, population, and the number of offices won't drive the average weekend ridership. It indicates that the average weekday ridership is mainly driven by commuting activities while the average weekend ridership is mainly induced by recreational activities such as shopping.

V.    CONCLUSIONS

Through backward stepwise regression procedure for variables selection, and the outlier test for determining the influential point, this paper recognized the influencing factors on Taipei metro station ridership of average weekday and average weekend, respectively. Influencing factors analyzed in this paper covered four dimensions: land use, social economic, accessibility and network structure. Different from previous work, we borrowed the conceptions, including degree centrality and betweenness centrality, from complex network theory to better quantify the network structure factors and related to the practical significance of metro networks. Also, the data used in this paper were easy to collect and had a certain theoretical basis. In order to improve the goodness-of-fit of the original regression model, we found Taipei Main Station was an influential point with much higher ridership than other stations through outlier test. Thus we added the dummy variable distinguishing transportation hub into the regression model. The regression performed much better than the previous models without considering influential points. The final models were simple to use with accessible data, and results significantly showed that the driving factors of ridership did differ from weekdays to weekends. In the regression models during different periods, the common driving factors were the land use of commerce, bus feeder system, distance to city center, days since the station opened and whether it was a transportation hub. Through comparing the influence factors on the ridership of different periods, it was noted that the ridership during weekdays was mainly affected by the commuting activities, while the ridership was driven by commercial access during weekends.

In terms of the implication of our study, the results can be used to understand the driving factors of metro travel demand at different times, thus provide a theoretical basis for traffic control and TOD planning. Firstly, TOD planning is suggested to be combined with metro network planning. The development around metro stations could be with high density and compact. Secondly, densely distributed offices nearby metro stations are key factors affecting commuting ridership, so the relevant strategies are necessary to be adopted to control traffic [23], plan TOD and balance commuting ridership. Thirdly, there is a strong interaction effect between the commercial development and daily trip ridership, which could be paid more attention to in TOD planning as it plays an important role in driving ridership. These findings can also inspire the metro planning and periphery development of other cities. Therefore, a further study could assess the factors influencing metro station ridership of other cities and compare the results.